\documentclass[aps,pra,twocolumn]{revtex4}
\usepackage{graphicx}
\usepackage{bm}
\usepackage{amsmath,amssymb,amsthm,dsfont,bm}
\usepackage{color}
\usepackage{wasysym}
\usepackage{ulem}
\usepackage{appendix}
\usepackage[dvipsnames]{xcolor}

\usepackage[english]{babel} 
\usepackage[utf8]{inputenc} 
\usepackage{hyperref}
\usepackage{cancel}

\begin{document}
\preprint{}
\title{Impact of quadrature measurement on quantum  coherence}
\author{Luc\'ia \'Alvarez and Alfredo Luis}
\email{alluis@fis.ucm.es}
\homepage{https://sites.google.com/ucm.es/alfredo/inicio}
\affiliation{Departamento de \'{O}ptica, Facultad de Ciencias
F\'{\i}sicas, Universidad Complutense, 28040 Madrid, Spain}
\date{\today}

\begin{abstract}
We examine the behavior of quadrature coherence under the measurement of the same field quadrature. This is  carried out with the help of a beam splitter, which implies the contribution of an auxiliary field state impinging at the other input port. To this end we consider the linear input-output transformation of a lossless beam splitter to relate input and output coherences, measured in terms of the $l_1$-norm. After obtaining a general input-output relation between coherences we apply the result to Gaussian and number states. For Gaussian states we obtain that coherence does not depend on the measurement outcome, and that the average coherence always equals the coherence of the reduced state, showing no average effect on coherence of the measurement. On the other hand, for number states the output coherence depends on the measurement, decreasing the relative coherence with increasing photon number. Finally, we consider relative-entropy as a measure of coherence to show that for number states and coherence measures other than the $l_1$-norm the average coherence no longer equals the coherence of the output reduced state.
\end{abstract}

\maketitle

\section{Introduction}

Coherence is an essential physical concept without which optics would be just geometrical optics and quantum mechanics would be a very complicated way of expressing classical mechanics. With the advent of quantum information, coherence has been rediscovered with a renewed interest in the form of resource theories for emerging quantum technologies \cite{JA06,BCP14,SAP17,CG19,AW16,WSRXCG21,SY22,SLWYYF17,H16,ATEMP22,BQSP15,BBCH16,CH15,SCLP19,JS23,RFWA18,BSFPW17,DKMS23,ZLYCW19,PRM15,BGW17,SL21}. In this situation the behavior of coherence under basic transformations and processes may be of much interest.  

In this regard, beam splitting is a well-known coherence maker in classical optics as a basic ingredient of most interferometric arrangements. This is so also in quantum optics, both in the Glauber-Sudarshan and Hilbert-space formalisms for coherence \cite{LS95,AL23b,DAL24,MW95,RG63a,RG63b,ECGS63}. A key feature is that its description, use, and properties are exactly parallel in classical and quantum optics.  

Most analyses on Hilbert-space coherence deal with discrete bases, such as photon-number. In this regard we have already examined the behavior of Hilbert-space coherence under beam splitting \cite{AL23b,DAL24}. In particular we have examined the optimal configuration of a set of beam splitters that maximize the quantum coherence of the output state. Such an analysis reveals the unsettling possibility of unlimited coherence growth with the increase of the number of beam splitters. 

Instead of the discrete, photon-number basis, here we focus on coherence in the continuous basis. More specifically the basis of eigenstates of a field quadrature \cite{AL22,AL23a,LA23}. This is interesting since continuous variables have a great impact as resources for quantum applications \cite{SLB05,ZSLF16,GF23,BR21,LL21,HKPU21}. For example, this is the case of the resource theories of squeezing and nonclassicality \cite{BY18,MI16}, as two basic practical resources for genuine quantum tasks. Moreover, the extension of quantum properties from finite-dimensional spaces to infinite dimension is not straightforward. This begins with the very definition of incoherent states, as examined in Ref. \cite{AL23a}, being the case that there are no physical incoherent states in a quadrature basis. Moreover, basic quantum-optical transformations, such as squeezing and beam splitting, are incoherent transformations that nevertheless can add coherence to partially coherent states.  There is also an intriguing infinite background of coherence present in the vacuum modes. So this issue is plenty of subtle points that provide a  suitable arena for a finer understanding of a physically relevant concept such it is quantum coherence.

In this regard, in Ref. \cite{AL23a} it was shown that the total quadrature coherence is conserved under beam splitting. In this work we add a further ingredient to this picture, which is measurement. If the behavior of coherence under transformations is interesting, both in the classical and quantum domains, the quantum realm provides another kind of evolution worth investigating under the form of state reduction due to measurement.  Intuitively, measurements should work as classical channels since under some circumstances transform quantum states into classical states \cite{KL22}. However, quantum measurements can present by themselves both nonclassicality and coherence, that may be considered also as a resource that may manifest in the measurement process \cite{YDXLS17,CGSSB19,GMSG21,XXTELYPZ20,BSLKN20}. In this work we consider that a quadrature measurement is performed in one of the outputs of the beam splitter. Then, we examine the coherence in the same quadrature basis of the emerging field state in the other output mode. That is to say, we are examining the behavior of quadrature coherence under a system transformation $\rho \rightarrow \rho^\prime$ induced by a quadrature measurement carried out with the help of a beam splitter. Beam splitting implies the contribution of an auxiliary field state $\rho_0$ impinging at the other input port (see Fig. 1).  

It is worth noting that both beam splitting and quadrature measurement are operations that exist exactly in the same terms in both classical and quantum optics. So we have a combination of transformation and measurement that are theory-agnostic so to speak. However, we are addressing here a fully quantum effect. This is because quantum beam splitting is always nontrivial due to vacuum fluctuations that enter through the normally unused input port. On the other hand, because of the entanglement produced by the beam splitter, state reduction by quantum measurement in one output port may affect the field leaving the other output port. These two issues are absent in a purely classical situation. Thus, this is a good scenario to study the behavior of quantum coherence.

\section{Coherence transformation}

The scheme is illustrated in Fig. 1. We start from a generic quantum state $\rho$ in a mode of the electromagnetic field described by the complex amplitude operator $a$. We mix it at a lossless ideal beam splitter with another input mode of complex amplitude $a_0$ in a fixed and known state $\rho_0$. At one of the outputs of the beam splitter we measure the quadrature $X^\prime_0$ of that output mode of the field with complex amplitude operator $a^\prime_0$. The output of the measurement will be denoted as $x^\prime_0$. The overall objective is to examine the coherence in the quadrature $X$ basis of the  state $\rho^\prime$ in the output mode with complex amplitude $a^\prime$, conditioned to the outcome $x^\prime_0$ of the measurement. We then compare the coherence of the so produced output state $\rho^\prime$ with the coherence of the input state $\rho$ in the same basis.
\begin{figure}[h]
    \includegraphics[width=5cm]{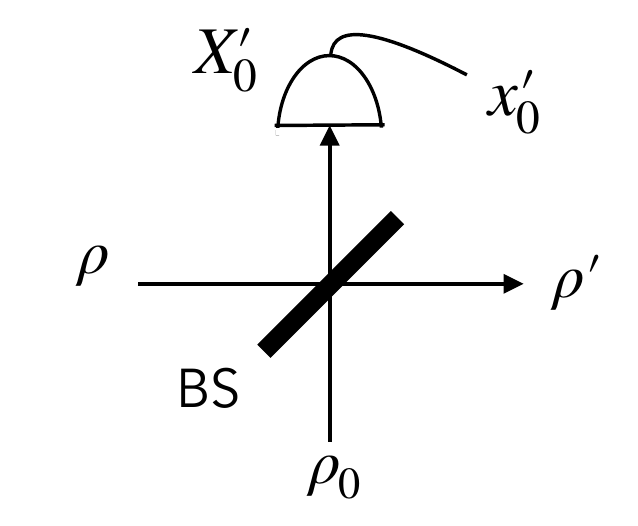}
    \caption{Scheme for the system transformation $\rho \rightarrow \rho^\prime$ induced by a measurement of quadrature $X^\prime_0$ with outcome $x_0^\prime$. The measurement is carried out with the help of a beam splitter BS that mixes the state $\rho$ of the signal mode with the field state $\rho_0$ in the auxiliary mode impinging at the other input port.}
    \label{scheme}
\end{figure}{}

As a suitable coherence measure, we utilize an extension to continuous quadratures of the $l_1$-norm of coherence \cite{BCP14,SAP17,AL22,AL23a}, that is 
\begin{equation}
    \mathcal{C} = \int_{-\infty}^\infty\int_{-\infty}^\infty dx dx^\prime | \langle x |\rho | x^\prime \rangle | ,
\end{equation}
where $|x\rangle$ are the eigenstates, not normalizable, of the quadrature $X$,
\begin{equation}
    X = \frac{1}{2} \left ( a^\dagger + a \right ) .
\end{equation}
For pure states $\rho = | \psi \rangle \langle \psi |$ this can be expressed in terms of the  quadrature wave-function $\psi (x)$ as
\begin{equation}
    \mathcal{C} = \left ( \int_{-\infty}^\infty |\psi (x) | \right )^2 ,  \quad  \psi (x) = \langle x |\psi \rangle .
\end{equation}

\bigskip

The field transformation induced by a lossless beam splitter can be described by the relation it establishes between the corresponding input and output complex amplitude operators 
\begin{equation}
    a^\prime = t a - r a_0, \quad a_0^\prime = t a_0 + r a ,
\end{equation}
that is exactly reproduced by the corresponding $X$ quadratures
\begin{equation}
\label{qt}
    X^\prime = t X - r X_0, \quad X_0^\prime = t X_0 + r X ,
\end{equation}
where $t,r$ are transmission and reflection coefficients assumed real without loss of generality, and satisfying the energy-conserving relation $t^2+r^2 = 1$. Throughout this work all quantities are dimensionless. 

\bigskip

In terms of field states, the action of the beam splitter can be described by a unitary operator $U$ such that the output state in modes $a^\prime, a^\prime_0$ is $U \rho \otimes \rho_0 U^\dagger$. This establishes the following relation between the input $| \widetilde{x_0} \rangle| x \rangle$ and output $| \widetilde{x_0^\prime} \rangle | x^\prime \rangle$ quadrature eigenstates, where for the sake of clarity we have marked with a tilde the eigenstates of the input and output quadratures in the auxiliary space, $X_0$ and $X^\prime_0$. That is
\begin{equation}
\label{Uiot}
    U | \widetilde{x_0} \rangle | x \rangle = |  \widetilde{x_0^\prime = t x_0 + r x} \rangle | x^\prime = t x - r x_0 \rangle .
\end{equation}
This transformation allows us to obtain a simple relation for the matrix elements in quadrature basis of the output state $\rho^\prime$ conditioned to the output of the measurement $x_0^\prime$, that is the quadrature matrix elements of the reduced state $\rho^\prime_u (x^\prime_0 )=\langle \widetilde{x_0^\prime} |\rho^\prime | \widetilde{x_0^\prime}\rangle$, where the subscript $u$ denotes that this is not normalized yet,
\begin{equation}
   \langle x |\rho^\prime_u (x^\prime_0 )| x^\prime \rangle = \langle x |\langle \widetilde{x_0^\prime} | U \rho \otimes \rho_0 U^\dagger | \widetilde{x_0^\prime} \rangle | x^\prime \rangle  .
\end{equation}
To compute these matrix elements we consider the inverse of transformations (\ref{qt}) and (\ref{Uiot}), that is 
\begin{equation}
    X = t X^\prime + r X^\prime_0, \quad X_0 = t X^\prime_0 - r X^\prime ,
\end{equation}
and
\begin{equation}
\label{iot}
    U^\dagger | \widetilde{x^\prime_0} \rangle| x^\prime \rangle = | \widetilde{t x^\prime_0 - r x^\prime }\rangle | t x^\prime + r x^\prime_0 \rangle .
\end{equation}
along with 
\begin{equation}
  \langle  x | \langle \widetilde{x^\prime_0} | U  = \langle \widetilde{t x^\prime_0 - r x }| \langle t x  + r x^\prime_0 |.
\end{equation}
Therefore
\begin{eqnarray}
\label{wio}
   &  \langle x |\rho^\prime_u (x^\prime_0 )| x^\prime \rangle = \langle x |\langle \widetilde{x_0^\prime} | U \rho \otimes \rho_0 U^\dagger | \widetilde{x_0^\prime} \rangle | x^\prime \rangle = & \nonumber \\
   & & \nonumber \\
   & \langle t x + r x^\prime_0 |\rho | t x^\prime + r x^\prime_0 \rangle \langle \widetilde{t x^\prime_0 - r x}|\rho_0 | \widetilde{t x^\prime_0 - r x^\prime} \rangle . &
\end{eqnarray}
This can be suitably normalized taking into account the probability $p(x^\prime_0)$ of the outcome $x^\prime_0$  
\begin{equation}
    p(x^\prime_0) = \int dx \langle x |\rho^\prime_u (x^\prime_0 )  | x \rangle ,
\end{equation}
so that the normalized output state is then
\begin{equation}
\label{una}
\rho^\prime (x^\prime_0)  = \frac{1}{p(x^\prime_0) } \rho^\prime_u (x^\prime_0). 
\end{equation}

A rather noticeable result is that the input-output transformation $\rho \rightarrow \rho^\prime$ is incoherent as it can be well appreciated in Eq. (\ref{wio}). Roughly speaking, if the input state $\rho$ tends to be diagonal in the quadrature basis, say $\langle x |\rho|x^\prime \rangle \rightarrow 0$ for $x \neq x^\prime$, so it will be the case of the conditioned output  $\langle x |\rho^\prime (x_0^\prime)|x^\prime \rangle \rightarrow 0$ \cite{BCP14,SAP17}. Despite this, the transformation may add coherence to partially coherent states, as we will clearly see next when considering particular examples. The fact that incoherent transformations may add coherence is a rather surprising and  interesting result that seemingly can only occur regarding coherence in continuous bases, not in discrete ones \cite{AL23a}.

\bigskip

The output coherence is 
\begin{equation}
\label{mc}
    \mathcal{C}^\prime (x^\prime_0) = \int_{-\infty}^\infty\int_{-\infty}^\infty dx dx^\prime | \langle x |\rho^\prime (x^\prime_0) | x^\prime \rangle | .
\end{equation}

Besides this single-shot output coherence, which may depend on the outcome of each measurement $x^\prime_0$, we may be interested in the average coherence $\overline{\mathcal{C}^\prime}$ by considering all possible results weighted by their probabilities
\begin{equation}
\label{aC}
\overline{\mathcal{C}^\prime} = \int d x^\prime_0 p(x^\prime_0) \mathcal{C}^\prime (x^\prime_0) =  \int d x^\prime_0 \mathcal{C}^\prime_u (x^\prime_0 ) ,
\end{equation}
where $\mathcal{C}^\prime_u (x^\prime_0 )$ denotes the coherence computed with the unnormalized state $\rho^\prime_u (x^\prime_0)$
\begin{equation}
\label{rC}
   \mathcal{C}^\prime_u (x^\prime_0 )= \int_{-\infty}^\infty\int_{-\infty}^\infty dx dx^\prime | \langle x |\rho^\prime_u (x^\prime_0 )| x^\prime \rangle | .
\end{equation}
This may be as well termed the nonreferring coherence in which we do not keep track of the result of the measurement. 

As a first result worth noting is that, maybe surprisingly, the averaged coherence $\overline{\mathcal{C}^\prime}$ actually does not depend on which measurement is carried out, or whether any measurement is actually carried out or not. This is because we can see in Eq. (\ref{aC}) that $\overline{\mathcal{C}^\prime}$ is equal to the coherence for the reduced state $\rho^\prime_r$ in the mode $a^\prime$, which is
\begin{equation}
    \rho^\prime_r = \int d x^\prime_0 \rho^\prime_u (x^\prime_0) ,
\end{equation}
since this operation is just taking the trace on the output mode $a^\prime_0$. Then, after Eqs. (\ref{aC}) and (\ref{rC})
\begin{equation}
\label{eq}
\overline{\mathcal{C}^\prime} = \mathcal{C}^\prime ( \rho^\prime_r ) .
\end{equation}
In general this is an effect of the coherence measure used, that is the $l_1$-norm. We will see that this is no longer the case if we were considering relative entropy as measure of coherence. We will illustrate this point later when examining the case of number states.

\bigskip

\section{Gaussian inputs}

For definiteness, and to obtain meaningful conclusions, let us consider the relevant case of Gaussian states, mixed in general, both in the system and auxiliary degrees of freedom. Without loss of generality we can consider centered states since coherence is invariant under quadrature shifts. Moreover, regarding coherence we can fully disregard imaginary parts \cite{SSM88}. Then, our expression for the quadrature matrix element of the input state is 
\begin{equation}
\langle x |\rho | x^\prime \rangle = \frac{1}{\sqrt{2 \pi \sigma^2}} \exp \left ( - \frac{x^2+x^{\prime 2}}{4 \sigma^2} \right ) \exp \left [ - \frac{(x- x^\prime)^2}{4 \mu^2} \right ] ,
\end{equation}
and equivalently for the auxiliary state with some other parameters $\sigma_0$ and $\mu_0$. These are the quantum counterparts of the Gaussian Schell model beams in classical optics \cite{MW95}. For these states there is a ready expression for the quadrature coherence as
\begin{equation}
     \mathcal{C} = 2 \sqrt{2 \pi} \frac{\sigma \mu}{\sqrt{2 \sigma^2 + \mu^2}} =  2 \sqrt{2 \pi}  \Delta  X \mathrm{tr} ( \rho^2) ,
\end{equation}
that depends both on the quadrature variance and the state purity 
\begin{equation}
    \Delta X = \sigma , \quad
   \mathrm{tr} ( \rho^2) = \frac{\mu}{\sqrt{2 \sigma^2 + \mu^2}}.
\end{equation}
The case of pure states may be reached in the limit $\mu/\sigma \rightarrow \infty$. It is worth noting that this Gaussian family of states includes thermal chaotic states with 
\begin{equation}
\sigma = \frac{\sqrt{2 \overline{n}+1}}{2} , \qquad \mu =  \frac{1}{2} \sqrt{\frac{2\overline{n}+1}{2\overline{n}(\overline{n}+1)}}  ,
\end{equation}
where $\overline{n}$ is the mean number of photons This can be quite interesting since thermal states tend to be quadrature incoherent states in the limit $\overline{n} \rightarrow \infty$, since 
\begin{equation}
\mathcal{C} = \sqrt{\frac{2 \pi}{2 \bar{n}+1} }.
\end{equation}
We also note that a rather unexpected relation emerges between field intensity and coherence, which is absent both in classical optics and in the Glauber-Sudarshan theory of quantum-optical coherence.
 
In this case, the output quadrature coherence $\mathcal{C}^\prime (x^\prime_0)$ for the reduced state in mode $a^\prime$ conditioned on the measurement result $x^\prime_0$ has a simple relation with the quadrature coherence $\mathcal{C}$ for the input state in mode $a$ and the quadrature coherence $\mathcal{C}_0$ of the auxiliary state $\rho_0$ in mode $a_0$. That is
\begin{equation}
\label{io0}
    \frac{1}{\mathcal{C}^{\prime 2} (x^\prime_0)} = \frac{t^2}{\mathcal{C}^2} + \frac{r^2}{\mathcal{C}_0^2} .
\end{equation}

This is a rather simple and meaningful relation. A relevant and surprising result to be noticed is that the coherence does not depend on the outcome of the measurement $x^\prime_0$, so we denote the output coherence simply by $\mathcal{C}^\prime$. Moreover, this further implies that the output coherence $\mathcal{C}^\prime$ is the same irrespective of whether we keep track of the measurement result, or even independent of whether the measurement is actually carried out or not. Therefore, the output coherence always equals the coherence of the reduced state,
\begin{equation}
   \mathcal{C}^{\prime } = \mathcal{C}^{\prime } (\rho^\prime_r) ,
\end{equation}
which agrees well with the general result  in Eq. (\ref{eq}) regarding averaged coherence. Although the coherence is independent of the result, the reduced state actually depends on $x^\prime_0$.

\bigskip

It can be easily seen that after Eq. (\ref{io0}) we have 
\begin{equation}
    {\rm max} \left ( \mathcal{C}, \mathcal{C}_0 \right ) \geq \mathcal{C}^{\prime } \geq  {\rm min} \left ( \mathcal{C}, \mathcal{C}_0 \right ). 
\end{equation}
In particular, we have that the system coherence increases whenever the auxiliary state $\rho_0$ is more coherent than the input system state $\rho$, that is 
\begin{equation}
 \mathcal{C}^\prime > \mathcal{C} \iff  \mathcal{C}_0  > \mathcal{C} ,
\end{equation}
irrespective of the beam splitter, provided that $r\neq 0$.

\bigskip

Two particular limits may be of interest in the general input-output relation (\ref{io0}). For example if $\mathcal{C}_0  \gg \mathcal{C}$ then we have  $\mathcal{C}^\prime \rightarrow \mathcal{C}/t$ as a kind of amplification of system coherence by a factor $1/t$. On the other limit if $\mathcal{C}_0  \ll \mathcal{C}$ and the auxiliary state tends to be incoherent, we have coherence loss tending to be locked at the coherence of the auxiliary state in the form $\mathcal{C}^\prime \rightarrow \mathcal{C}_0/r$. 

This behavior agrees well with the incoherent nature of the transformation, discussed above in the paragraph between Eqs. (\ref{una}) and (\ref{mc}). In the limit of incoherent input states $\mathcal{C} \rightarrow 0$ we have incoherent output states, $\mathcal{C}^\prime \rightarrow 0$. Nevertheless, the transformation may add coherence to partial coherent states $\mathcal{C} \neq 0$. Equivalently, if the auxiliary state is incoherent $\mathcal{C}_0 \rightarrow 0$ the output state will also tend to be incoherent. 

\bigskip

The variation of coherence in Eq. (\ref{io0}) can be well illustrated by Fig. 2 where we represent $\mathcal{C}^\prime-\mathcal{C}$ as a function of the transmission coefficient $t$ for vacuum in the auxiliary mode and a pure squeezed state in the signal mode. We consider two cases, with quadrature uncertainties above $\Delta X =1$ (dashed line) and below $\Delta X =1/4$ (solid line) the vacuum level $\Delta X =1/2$.
\begin{figure}[h]
    \includegraphics[width=7cm]{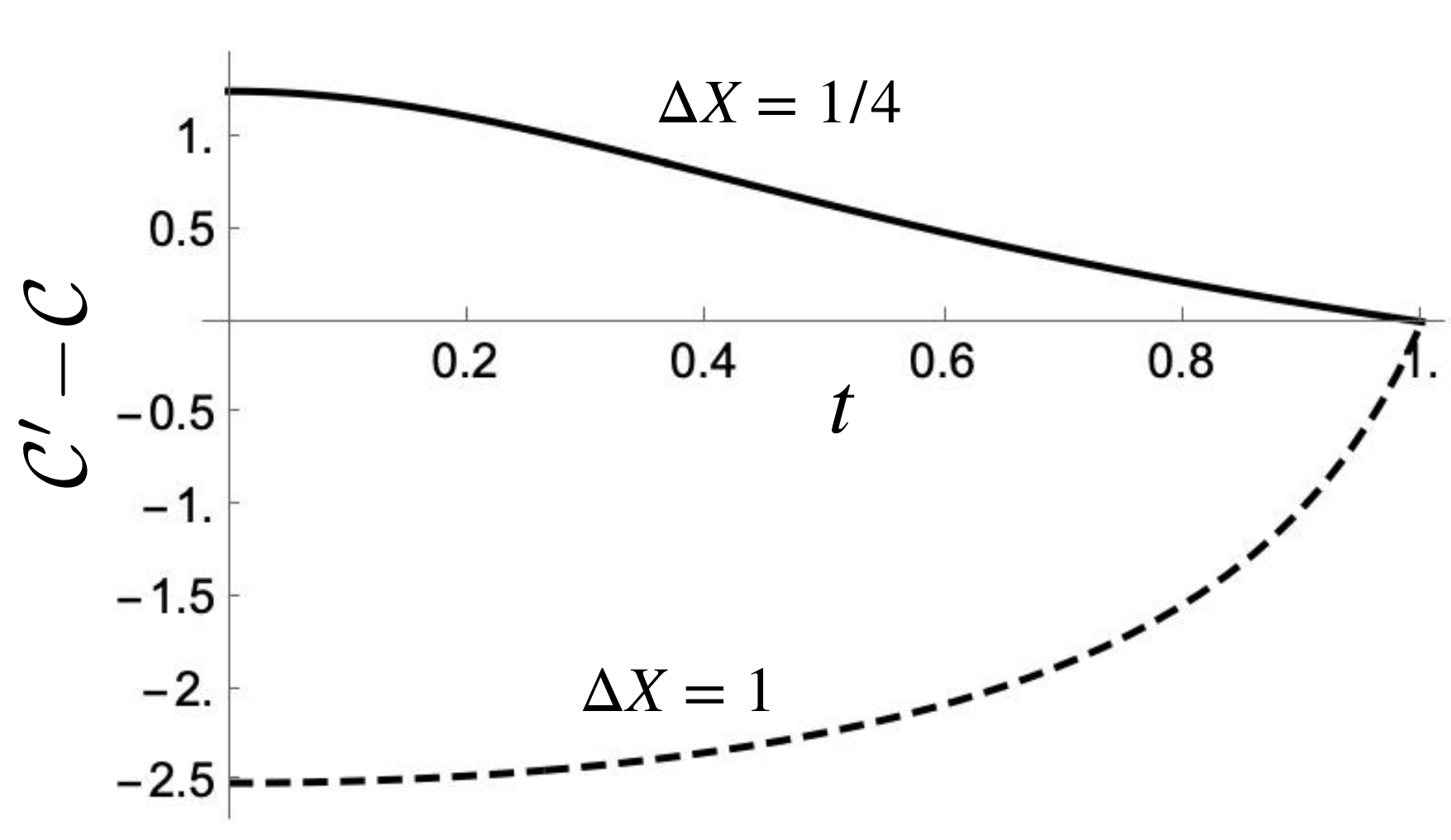}
    \caption{Transformation of coherence expressed in the form $\mathcal{C}^\prime-\mathcal{C}$, where $\mathcal{C}$ and $\mathcal{C}^\prime$ are the input and output coherence, respectively. This is plotted as a function of the transmission coefficient $t$ for the vacuum state in the auxiliary mode. The state in the signal mode is pure and squeezed, with quadrature uncertainties $\Delta X =1/4$ (solid line) and $\Delta X =1$ (dashed line).}
\end{figure}{}

\bigskip

Finally, we can notice an interesting related result that applies to pure Gaussian states that are also minimum uncertainty states $\Delta X \Delta Y = 1/4$, where quadrature $Y$ is
\begin{equation}
    Y = \frac{i}{2} \left ( a^\dagger - a \right ) .
\end{equation}
For these states $\Delta X \Delta Y = 1/4$  means that $\mathcal{C}_X \mathcal{C}_Y= 2\pi$, where here the subscripts on $\mathcal{C}$ indicate the quadrature basis where coherence is computed. With this we can easily translate Eq. (\ref{io0}) into a relation between input and output coherence in the quadrature $Y$ after the measurement of the quadrature $X$, that is 
\begin{equation}
     \mathcal{C}_Y^{\prime 2} =  t^2 \mathcal{C}_Y^2 + r^2 \mathcal{C}_{Y0}^{2}.
\end{equation}

\bigskip

\section{Number-state input}

As a further example let us consider photon-number states for the system input mode. The auxiliary mode will be always in the vacuum state. For simplicity we always consider a 50 \% beam splitter. The number states read in the quadrature $X$ representation as follows
\begin{equation}
    \psi_n (x) = \sqrt{\frac{\sqrt{2/\pi}}{2^n n!}}H_n(\sqrt{2} x)e^{-x^2} ,
\end{equation}
where $H_n$ are the Hermite polynomials. Quadrature coherence increases with the number $n$ as shown in Fig. 3, maybe because $\Delta X$ increases for increasing $n$. Here again we can note an emerging relation between field intensity and coherence, in this case increasing coherence with increasing intensity, contrary to the Gaussian example above.

\begin{figure}[h]
    \includegraphics[width=7cm]{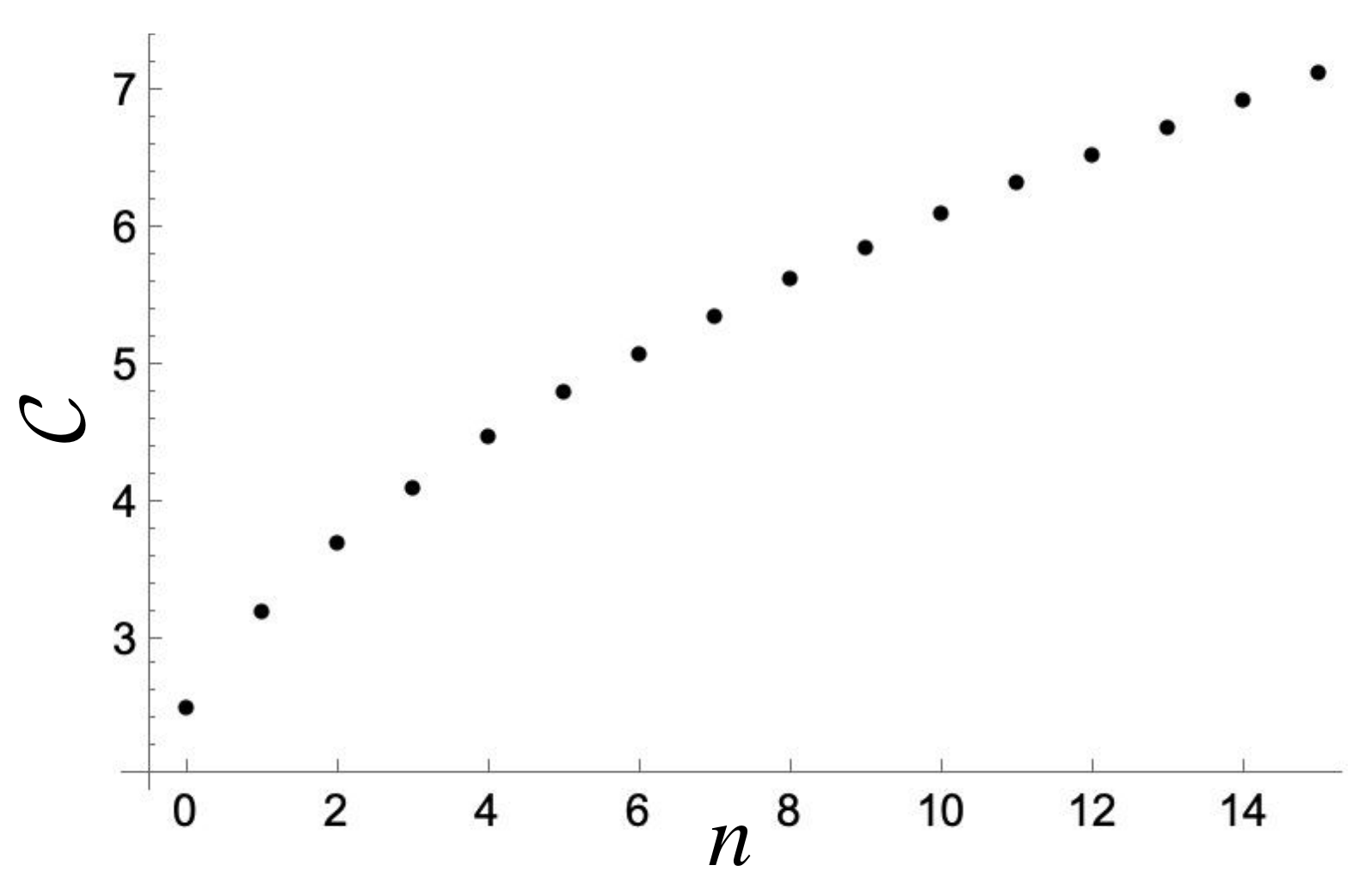}
    \caption{Plot of the quadrature coherence $\mathcal{C}$ for number states as a function of the number of photons $n$.}
\end{figure}{}

In this case, we get the expected result that the coherence of the output state will depend on the result of the quadrature measurement $x^\prime_0$ as shown in Fig. 4 where we plot the output coherence relative to the input one $\mathcal{C}^\prime (x^\prime_0)/\mathcal{C}$ as a function of $x^\prime_0$ for $n=1$ (solid line), $n=2$ (dashed line), and $n=3$ (dotted line). 

\begin{figure}[h]
    \includegraphics[width=7cm]{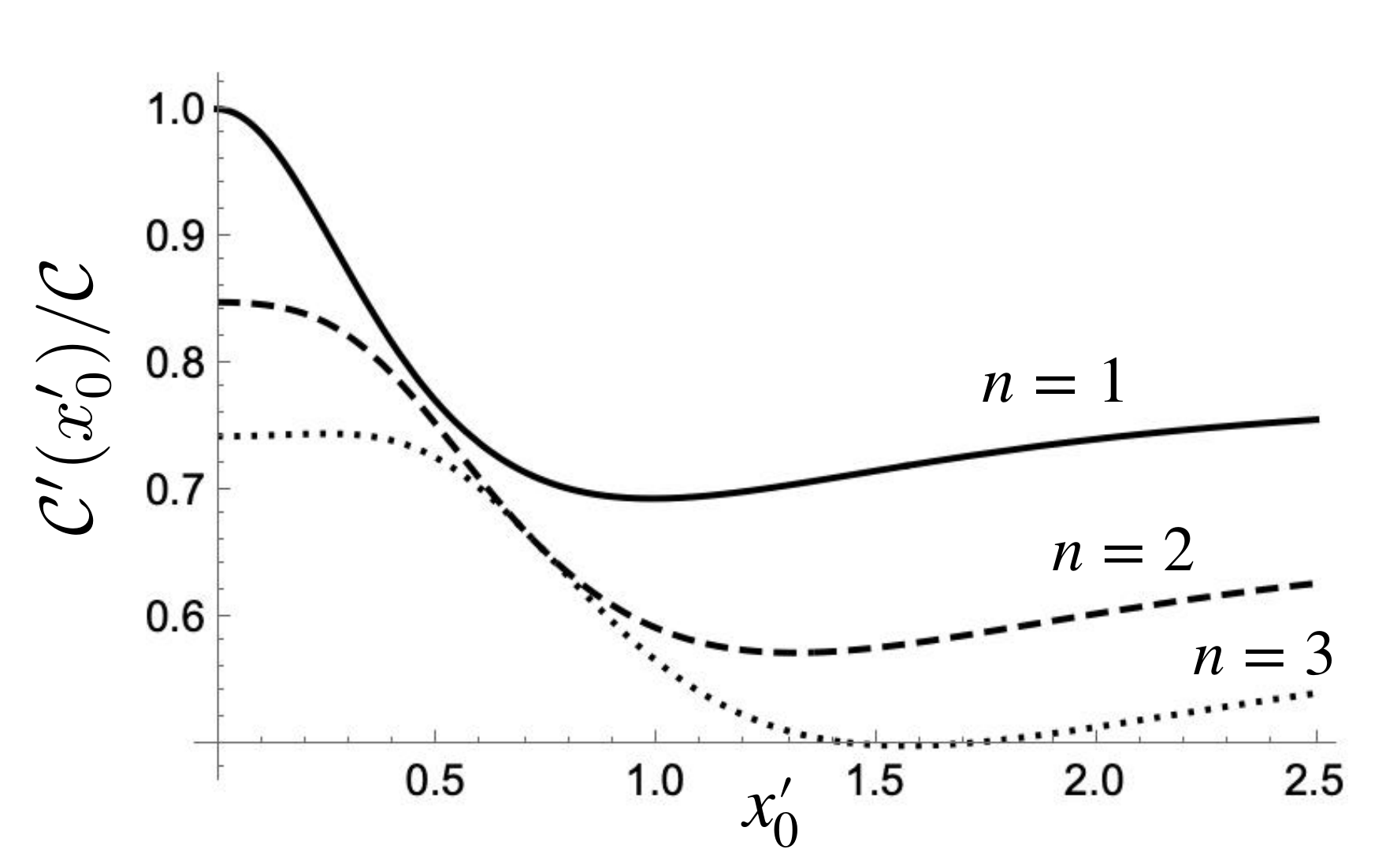}
    \caption{Plot of the output coherence $\mathcal{C}^\prime (x^\prime_0)$ relative to the input coherence $\mathcal{C}$ as a function of the outcome $x^\prime_0$ of the quadrature measurement. This is plotted for input number states with different photon numbers $n$, that is for $n=1$ (solid line), $n=2$ (dashed line), and $n=3$ (dotted line).}
\end{figure}{}

The probabilities $p(x^\prime_0)$ of the corresponding outcomes $x^\prime_0$ are plotted in Fig. 5 as a function of $x^\prime_0$ for $n=1$ (solid line), $n=2$ (dashed line), and $n=3$ (dotted line). We consider just positive values for $x^\prime_0$ since curves are symmetric. These probabilities  mean that outcomes $x_0^\prime$ above some threshold, say $x^\prime_0 >2$, are meaningless so a more meaningful approach is to consider relative coherence weighted by the probability $p(x^\prime_0)$, this is to plot $p(x^\prime_0) \mathcal{C}^\prime (x^\prime_0 )/\mathcal{C}$. This is plotted in Fig. 6.

\begin{figure}[h]
    \includegraphics[width=7cm]{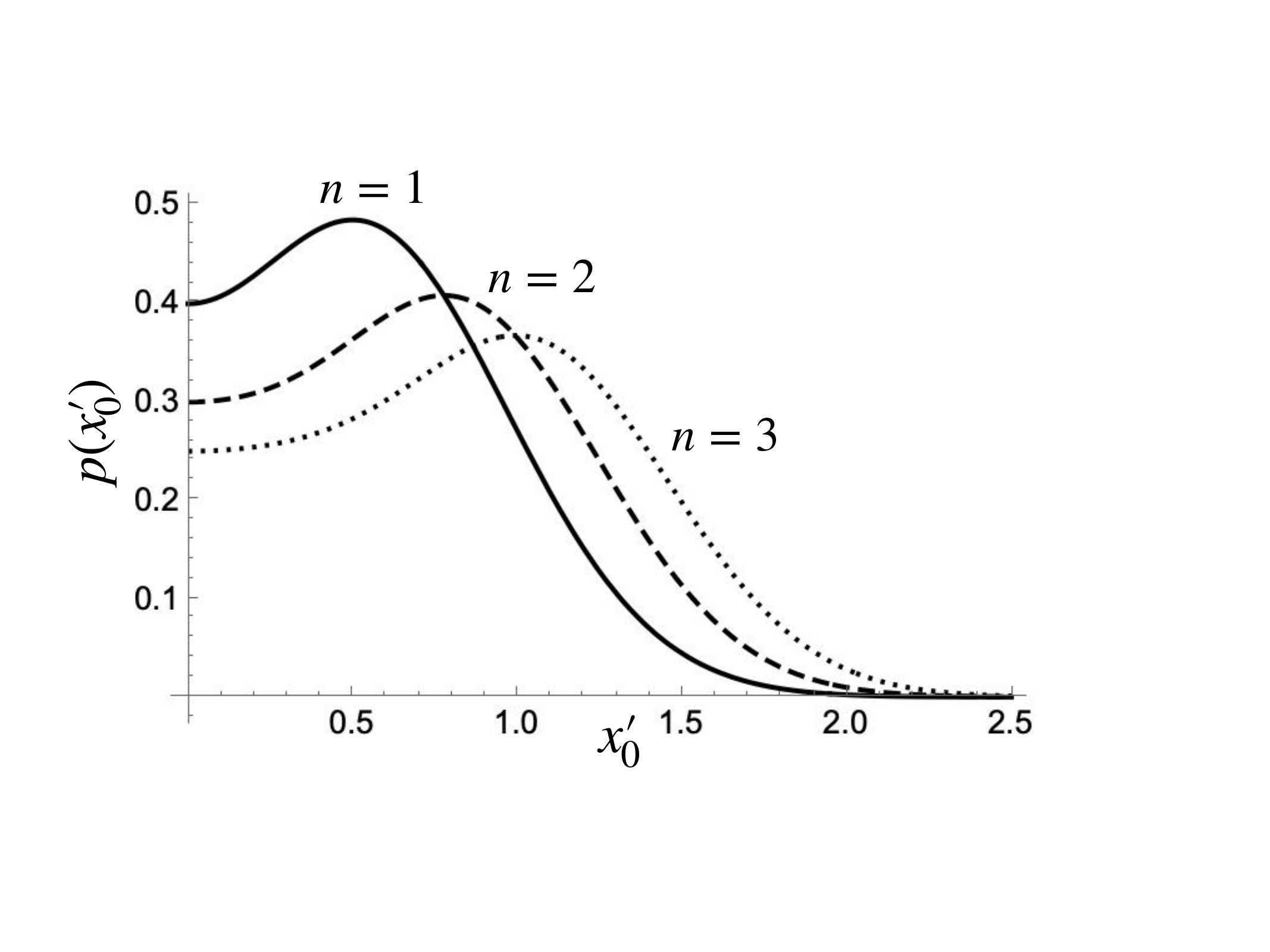}
    \caption{Plot of the probability $p(x^\prime_0)$ for the outcome $x^\prime_0$ of the quadrature measurement as a function of the outcome $x^\prime_0$ of the quadrature measurement for input number states with different photon numbers $n$, that is  $n=1$ (solid line), $n=2$ (dashed line), and $n=3$ (dotted line).}
\end{figure}{}

\begin{figure}[h]
    \includegraphics[width=7cm]{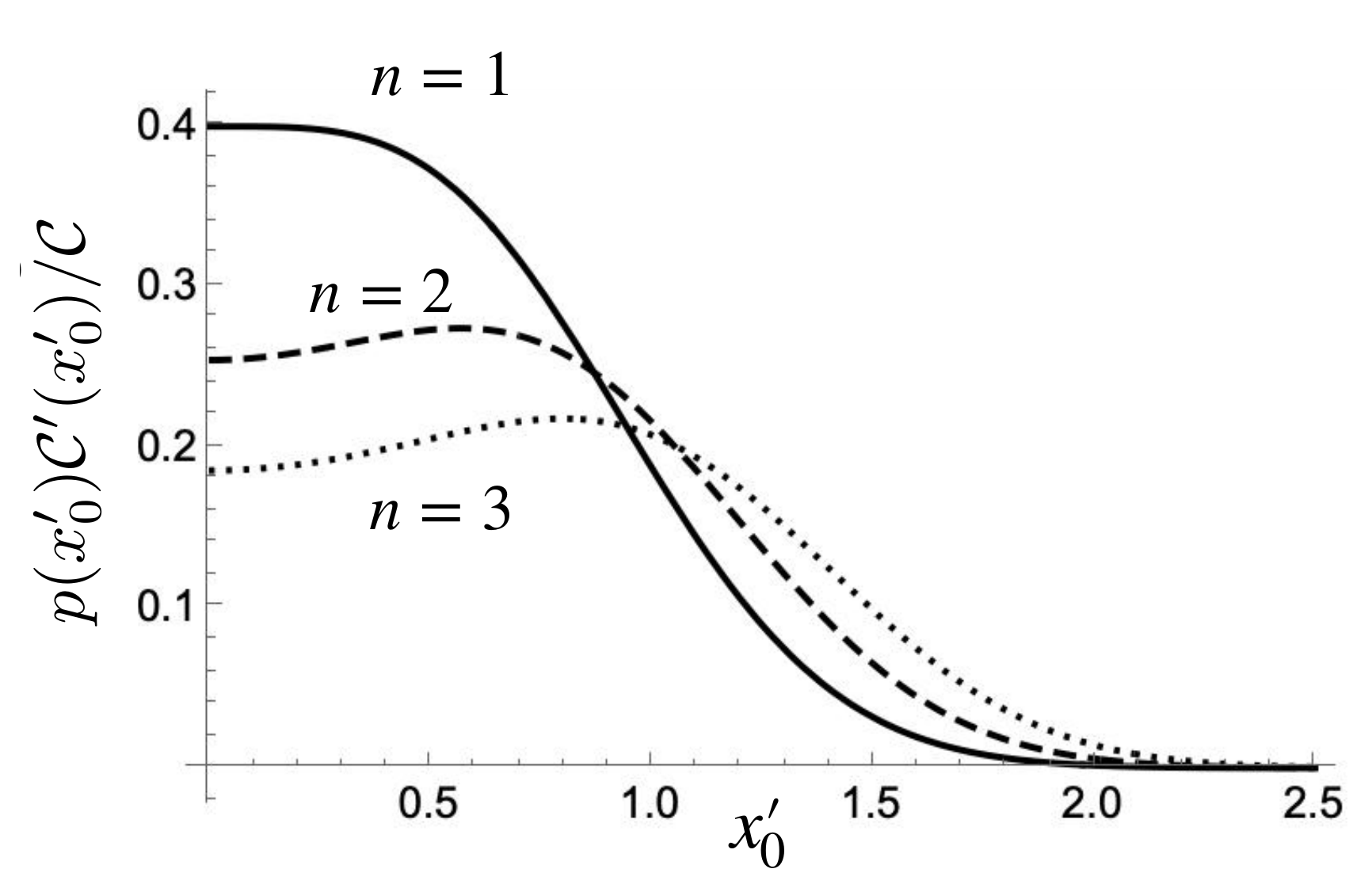}
    \caption{Plot of the output coherence $\mathcal{C}^\prime (x^\prime_0 )$ relative to the input one $\mathcal{C}$ weighted by the probability $p(x^\prime_0)$ as a function of the outcome $x^\prime_0$ of the quadrature measurement. This is plotted for input number states with different photon numbers $n$, that is for $n=1$ (solid line), $n=2$ (dashed line), and $n=3$ (dotted line).}
\end{figure}{}

Regarding the absolute averaged coherence we get the behavior represented in Fig. 7, while in Fig. 8 we plot the relative averaged coherence, both as functions of the input number of photons $n$. In agreement with the outcome-dependent results in Fig. 4, it can be well appreciated in Fig. 8 that the the averaged output coherence  $\overline{\mathcal{C}^\prime}$ is always less than the input coherence $\mathcal{C}$, and that the coherence loss increases with increasing number of input photons.

\begin{figure}[h]
    \includegraphics[width=7cm]{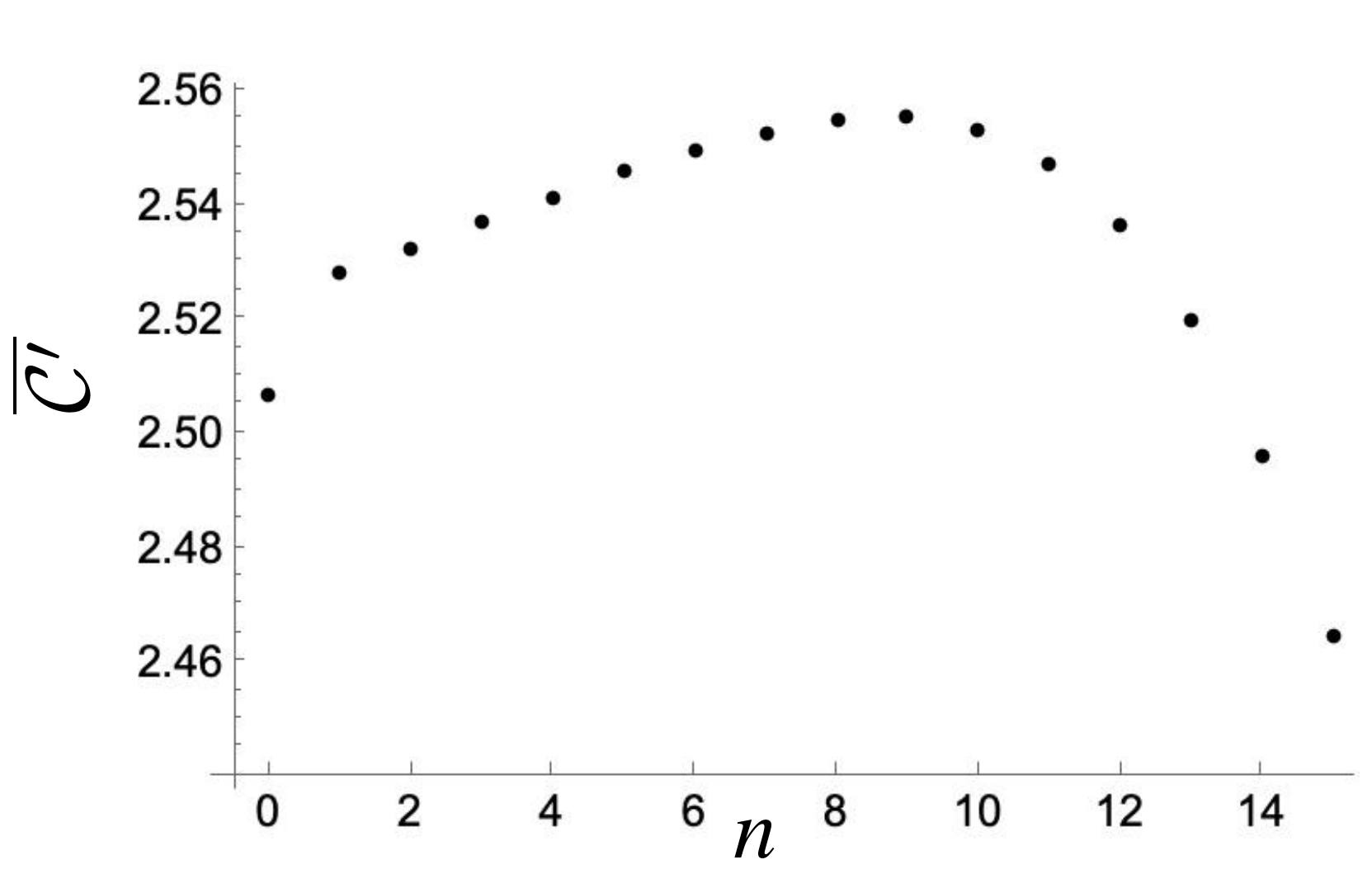}
    \caption{Plot of the averaged output coherence $\overline{\mathcal{C}^\prime}$ for input number states as a function of the number of photons $n$.}
\end{figure}{}

\begin{figure}[h]
    \includegraphics[width=7cm]{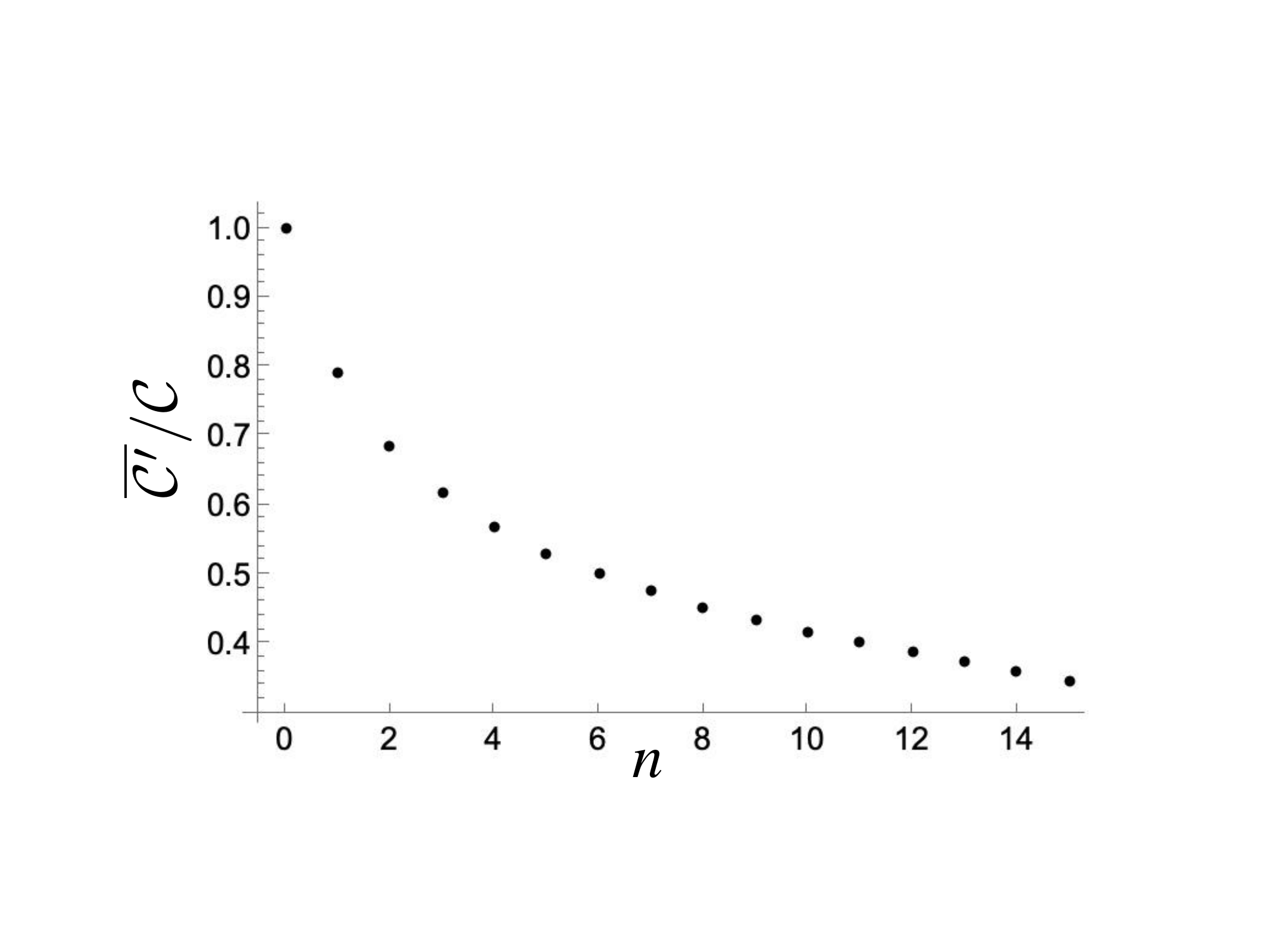}
    \caption{Plot of the averaged output coherence $\overline{\mathcal{C}^\prime}$ relative to the coherence of the input state $\mathcal{C}$ for input number states as a function of the number of photons $n$.}
\end{figure}{}

\bigskip

\section{Average versus reduced-state coherences via relative entropy}

We showed above that for the $l_1$-norm we have the universal exact equality between the average coherence $\overline{\mathcal{C}^\prime}$ and the coherence $\mathcal{C}^\prime (\rho^\prime_r)$ for the reduced state $ \rho^\prime_r$, that is $\overline{\mathcal{C}^\prime} = \mathcal{C}^\prime (\rho^\prime_r)$. Let us show that this is no longer the case if we use a different measure of coherence, such as the relative entropy of coherence for instance
\begin{equation}
    S(\rho||\rho_d) = \mathrm{tr} \left ( \rho \ln \rho \right ) - \mathrm{tr} \left ( \rho \ln \rho_d \right ) ,
\end{equation}
where $\rho_d$ is the diagonal part of $\rho$ in the corresponding basis, its incoherent part so to speak. In the appropriate limits we get \cite{AL23a}
\begin{equation}
\label{re}
   S(\rho||\rho_d) = \mathrm{tr} \left ( \rho \ln \rho \right ) - \int_{-\infty}^\infty dx p(x) \ln p(x) ,
\end{equation}
where $p(x)$ is the quadrature probability distribution in the state $\rho$. For the sake of simplicity we consider just the case of a one-photon state as system input and vacuum as input auxiliary state, and a beam splitter of variable transmittance and reflectance. The reduced state after every measurement is pure so the first factor in the relative entropy is absent, while the quadrature probability is 
\begin{equation}
p(x) = \frac{1}{p(x^\prime_0)} \left ( t\psi_1 (x) \psi_0 (x^\prime_0 )+  r\psi_0 (x) \psi_1 (x^\prime_0 ) \right )^2 ,
\end{equation}
where $p (x^\prime_0 )$ is the probability of the outcome $x^\prime_0$
\begin{equation}
p(x^\prime_0) = t^2 \psi_0^2 (x^\prime_0)+ r^2 \psi_1^2 (x^\prime_0) . 
\end{equation}
On the other hand, the reduced state $\rho^\prime_r$ is mixed, being diagonal in the photon-number basis
\begin{equation}
\rho^\prime_r = r^2 |0\rangle \langle 0 | + t^2 |1\rangle \langle 1 |,
\end{equation}
leading to a quadrature probability distribution 
\begin{equation}
p_r(x) = r^2 \psi_0^2 (x)+ t^2 \psi_1^2 (x) . 
\end{equation}
With this we have all the ingredients to compute both the average coherence and the coherence of the reduced state. We plot them in Fig. 9 as functions of the transmission coefficient $t$, showing that they are different, except in the trivial cases $t=0,1$. This is a rather interesting effect of measurement on coherence. As we commented in the Introduction, this is a purely quantum effect. Actually the net effect is positive, in the sense that the output average coherence after measurement  $\overline{\mathcal{C}^\prime}$ is above the coherence of the reduced state $\mathcal{C}^\prime_r$, which is the coherence one might expect from a purely classical-like reasoning.

\begin{figure}[h]
    \includegraphics[width=7cm]{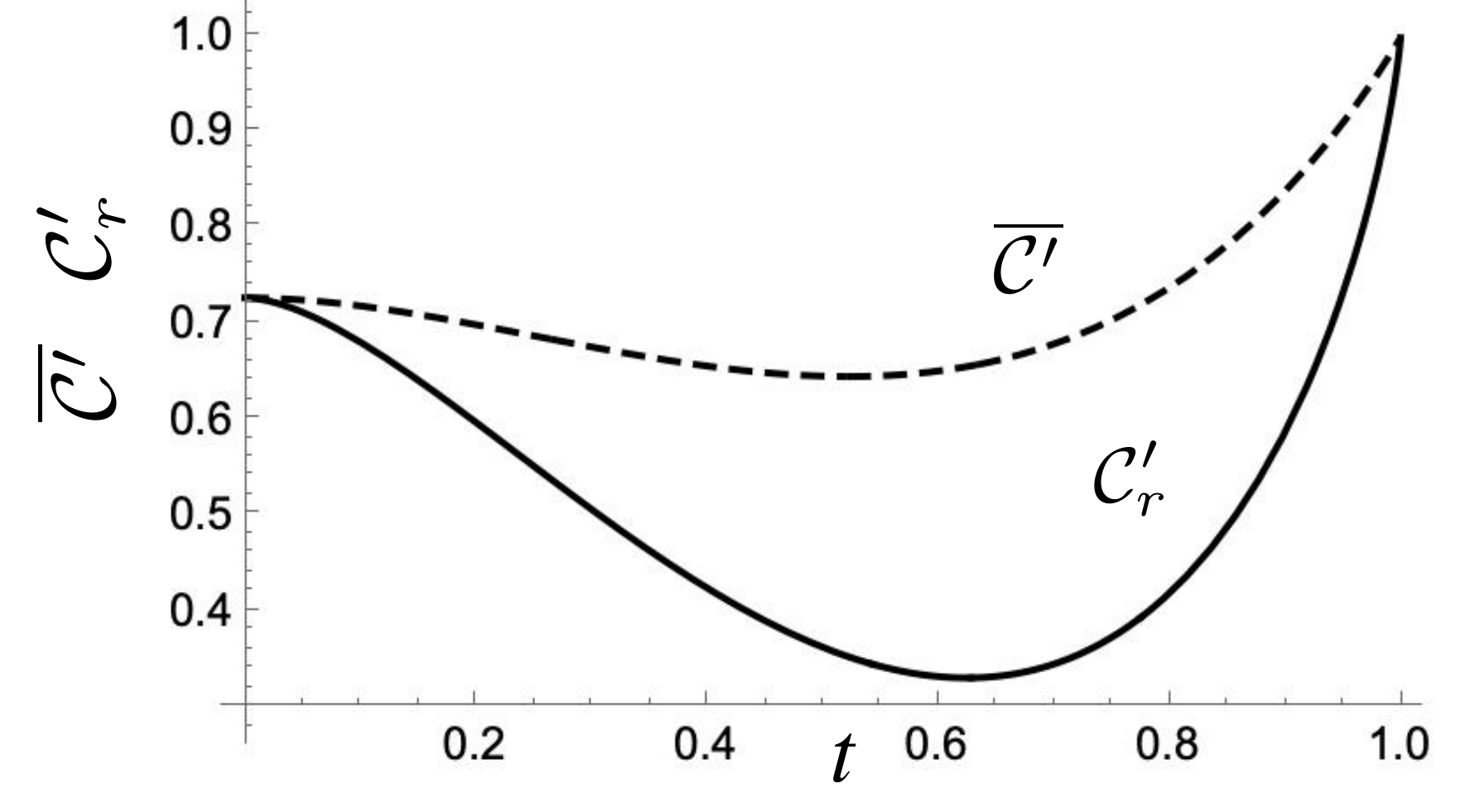}
    \caption{Plots of the output average coherence $\overline{\mathcal{C}^\prime}$ (dashed line) and the coherence for the reduced state $ \rho^\prime_r$ in the signal mode $\mathcal{C}^\prime_r =\mathcal{C}^\prime (\rho^\prime_r)$ (solid line) as a function of the transmission coefficient $t$ for a one-photon input state.}
\end{figure}{}

\bigskip

\section{Conclusions}

We have examined the impact of quadrature coherence under measurement of the same field quadrature. We may note first that this is a fully quantum effect. In a classical scenario the observation of the field state by means of a beam splitter does not require the presence of any auxiliary state, say that the classical vacuum equals nothing, and the measurement in one of the output modes has no effect on the other. In such a case the output field state is just the input one multiplied by the transmission coefficient, the measurement having no effect at all. So, the quantum effect is a combination of interference and state reduction.

We have derived a suitable relation between input and output coherences for Gaussian states that properly accounts for the physical processes taking place. There is the interesting result that coherence does not depend on the measurement outcome. In this regard, it is also worth noting that for every state, whenever $l_1$-norm is used, the average coherence always equals the coherence of the reduced state, showing no average effect on coherence of the measurement.  

Things are different for input number states. On the one hand, the output coherence depends on the measurement outcome. On the other hand, we have used the relative entropy as measure of coherence to show that the average coherence no longer equals the coherence of the reduced state, showing that the measurement actually increases the amount of coherence. 

We believe these results may be useful for a better understanding of the key concept of coherence in its many interesting manifestations. 

\section*{ACKNOWLEDGMENTS} 
We thank Dr. L. Ares for her interest and valuable comments.

\section*{DATA AVAILABILITY}
The data that support the findings of this article are not
publicly available. The data are available from the authors
upon reasonable request.

\end{document}